# Persistence length of chromatin determines origin spacing in *Xenopus* early-embryo DNA replication: Quantitative comparisons between theory and experiment

Suckjoon Jun [1*], John Herrick [2*], Aaron Bensimon [2*], and John Bechhoefer [1*]

[1] Department of Physics, Simon Fraser University, Burnaby, B.C., V5A 1S6, Canada
[2] Unité de Stabilité des Génomes, Département de Structure et Dynamique des Génomes, Institut Pasteur, 25-28, rue du Dr. Roux, 75724, Paris Cedex 15, France

*Corresponding authors:
John Bechhoefer (johnb@sfu.ca)
ph: 604-291-5924 / fax: 604-291-3592

and

Aaron Bensimon (abensim@pasteur.fr)
ph: +33 (0)1 40 61 36 89 / fax: +33 (0)1 45 68 87 90

## Abstract

In *Xenopus* early embryos, replication origins neither require specific DNA sequences nor is there an efficient S/M checkpoint, even though the whole genome (3 billion bases) is completely duplicated within 10-20 minutes. This leads to the "random-completion problem" of DNA replication in embryos, where one needs to find a mechanism that ensures complete, faithful, timely reproduction of the genome without any sequence dependence of replication origins. We analyze recent DNA replication data in *Xenopus laevis* egg extracts and find discrepancies with models where replication origins are distributed independently of chromatin structure. Motivated by these discrepancies, we have investigated the role that chromatin looping may play in DNA replication. We find that the loop-size distribution predicted from a wormlike-chain model of chromatin can account for the spatial distribution of replication origins in this system quantitatively. Together with earlier findings of increasing frequency of origin firings, our results can explain the random-completion problem. The agreement between experimental data (molecular combing) and theoretical predictions suggests that the intrinsic stiffness of chromatin loops plays a fundamental biological role in DNA replication in early-embryo *Xenopus* in regulating the origin spacing.

# Introduction

In prokaryotes such as *E. coli*, in simple eukaryotes such as *S. cerevsiae*, and in somatic cells, genome sequence plays an important role in defining origins of DNA replication.[1] In *Xenopus* and *Drosophila* early embryos, by contrast, replication origins do not require any specific DNA sequences. If potential origins are distributed randomly along the genome, one expects a geometric (exponential) distribution of separations. Because the length of S phase is determined by the replication of the *entire* genome, even relatively rare long gaps could prolong S phase beyond its observed duration of 10-20 minutes for complete duplication of the whole genome (3 billion bases).[2,3] The problem is all the more acute in that early embryo cells lack an efficient S/M checkpoint,[4] which is used by many eukaryotic cells to delay entry into mitosis in the presence of unreplicated DNA. This problem is formally stated as the "random-completion problem,"[5] and, because of the reasons explained above, its solution requires a mechanism that regulates replication other than sequence.

Roughly, two approaches have been advanced to resolve the random-completion problem:[6] In the first scenario ("origin redundancy" model), potential origins exist in abundance and initiate stochastically throughout S phase. This allows large gaps to be "filled in" during the later stages of S phase.[7,8] In the second scenario ("fixed spacing" model), one postulates a mechanism that imposes regularity in the distribution of potential origins, thus preventing the formation of problematic large gaps between origins.[9] In this article, we shall show that consideration of recent experimental results on early embryo *Xenopus* replication leads to a more nuanced, "intermediate" view that incorporates elements of both scenarios and, more important, suggests a biological picture in which the secondary structure of chromatin – looping in particular – plays an important biological role in DNA replication.

One recent development is that new experimental techniques now make it possible to extract large amounts of data from the replication process. For example, molecular-combing[10] and direct visual hybridization (DIRVISH)[9,11] techniques can give detailed statistics about numbers and sizes of replicated domains as averaged over the genome, as well as many other related quantities.[11,12] Alternatively, gene-chips have given information about how replication proceeds at specific locations of the genome.[13,14] Finally, the recently demonstrated "Chromosome Conformation Capture" (3C) technique gives information about average dynamical configurations of chromosomes, using crosslinking to measure interaction frequencies between different genetic loci.[15] The amount and quality of the data from these recent experiments is stimulating the formation of quantitative models of DNA replication.[8]

Here, we show that recently obtained molecular-combing data on DNA replication in early-embryo *Xenopus laevis* are most naturally explained by postulating that chromatin forms loops at "replication factories"[16,17] and that these loops control origin spacing (Fig. 1); It is important to note that the size of such a loop is not arbitrary. The stiffness of the polymer means that loops that are too small cost too much energy. If a loop is too large, there will be too many conformations to explore for the ends to meet, and it thus costs too much entropy. Balancing these effects gives an optimal loop size,[18] which leads to an origin-exclusion zone, since origins are connected by at least a single loop.

The sizes of the postulated loops extracted by fitting to experimental data turn out to be comparable to those obtained independently in single-molecule measurements of chromatin stiffness in other systems.[15,19] Because the size of a polymer loop is controlled by its stiffness, we can link the physical properties of chromatin, when considered as a semiflexible polymer, to origin spacing during DNA replication. As we shall see, the physical properties of chromatin loops can explain both the observed regularity of initiation spacings[9] and the existence of an "origin-exclusion zone,"[7] where origin firing is inhibited, reconciling apparently contradictory views on the nature of the mechanism that ensures rapid and complete genome replication in early embryos. Although our results concern one particular system, there is reason to suspect that they may apply more generally.

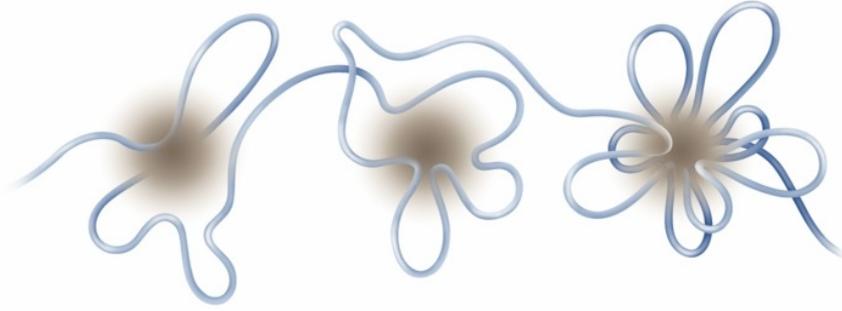

**Figure 1 - Replication factory and chromatin loops.** Schematic description of how chromatin folding can lead to replication factory with loops. The loop sizes are not arbitrary (see Materials and Methods).

## Materials and Methods

### Analysis of Molecular Combing Experiments on Early-Embryo *Xenopus*.

We analyzed data from the recent molecular combing experiment by Herrick *et al.*[12] The data are available on request. These experiments used *Xenopus* sperm nuclei in *Xenopus* egg extracts and two-color fluorescent labeling of DNA bases to study the kinetics of DNA replication in this system. One begins by labeling the sperm chromatin with a single fluorescent dye (biotin-dUTP, visualized as Texas Red). At some time point $t'$ during the replication process, one adds the second dye (dig-dUTP, green) and thus DNA replicated after $t'$ are labeled with two colors (predominantly green). Fully replicated DNA are stretched out *uniformly* on a glass surface using molecular combing and examined under a microscope (stretching factor: $1\mu m = 2.0 \pm 0.1$ kb; see Fig. 8 in Ref. 20). The alternating red-and-green regions form a snapshot of the replication state of the DNA fragment at the time the second dye was added. Varying that time in different runs allows one to systematically look at the progression of replication throughout S phase.

In previous analysis, we examined the average lengths of eyes and holes at different times during S phase.[8] Here, we focused on the distribution $\rho_{i2i}$ of eye-to-eye distances in order to test the origin spacings predicted by the wormlike chain model of chromatin fibers, as well as the origin synchrony.

We also generalized the correlation measurements of Blow *et al.*[9] In our simulations, we can detect origin synchrony through correlations in the sizes of nearby replicated domains (or eye sizes). Adjacent (small) eyes of similar size will have initiated at about the same time. The correlation coefficient is defined as

$$C(|i-j|) \equiv \frac{\left\langle (s_i - \langle s_i \rangle)(s_j - \langle s_j \rangle) \right\rangle}{\sqrt{\left\langle (s_i - \langle s_i \rangle)^2 \right\rangle \left\langle (s_j - \langle s_j \rangle)^2 \right\rangle}},$$

where $s_i$ ($s_j$) is the *i*-th (*j*-th) eye size and brackets (<...>) denote average values. The neighborhood distance |*i-j*| indicates how far

two eyes are apart. For example, $C(1)$ is the correlation coefficient for nearest neighbors, $C(2)$ for next-nearest, and so on.

## Looping of a Helical, Wormlike Polymer Chain: Statistics and Dynamics.

In forming loops (see Fig. 1, for example), polymers that have an intrinsic stiffness such as chromatin cannot have arbitrary loop sizes. The optimal loop size is 3-4 times the persistence length (a measure of the polymer stiffness).[18] Previous work dealing with looping in biological contexts has implicitly assumed that looping is a reaction-limited process, i.e., one where the reactive groups meet many times before actually binding.[21] In this limit, the kinetic distribution of loop sizes is identical to the distribution of loops in thermal equilibrium. For this case, Shimada and Yamakawa (SY) derived an approximate expression, valid for $l < 10 \, l_p$:[22]

$$(1) \quad G(l) = 896.32 \cdot l^{-5} \cdot \exp\left(-\frac{14.054}{l} + 0.246 \cdot l\right).$$

Here, $G(l) \cdot dl$ is the probability for finding a loop whose size is between $l$ and $l+dl$, where $l = L / l_p$, with $L$ the contour length of the polymer and $l_p$ the persistence length. Notice that for small $l$, the loop-formation probability is exponentially suppressed, which provides a natural explanation for an origin-initiation exclusion zone. The peak of the SY distribution at $l = 3.4$ can be expected to correspond to enhanced initiations. Finally, for $l \geq 10$, the probability decreases rapidly, which makes the formation of single large chromatin loops unlikely. Note again that Eq. 1 does not accurately describe this large-$l$ limit, which has been modeled more accurately as a Gaussian chain.[23]

If the dynamics are diffusive, i.e., if the reactive groups bind the first time they encounter each other within some small reaction range $\alpha$ ($< 1$), we can show that the SY approximation continues to hold in the regime where the loop-size is less than a few times the persistence length, and the loop-formation time $\tau_c$ is given by

$$(2) \quad \tau_c(l) = C \frac{1}{\alpha D} \frac{l_p^2}{G(l)},$$

where $C$ is a dimensionless prefactor that is practically a constant ($\sim 10^{-1}$) for all $l$, and $D$ is the diffusion constant.[24] This "first return time" $\tau_c$ predicted by Eq. 2 is very short ($10^{-3}$ to $10^{-2}$ seconds for chromatin, comparable to that of linear dsDNA[25]), implying that loop-formation dynamics are much faster than replication time scales ($\sim$ 20 minutes).

Finally, one further approximation that has been made in this and previous work on looping is that the reactive groups are assumed to be the polymer ends, whereas in the case of chromatin, origins along the DNA (i.e., not at the ends of the DNA) are assumed to bind to replication factories that have already bound a neighboring origin, which is also in general not at the end of the chromatin molecule. We believe that this is unlikely to be an important complication.

Note that while the loop-size distribution does not accurately follow the SY distribution outside the so-called Kramers regime where Eq. 2 was derived, the folding of chromatin falls within this limit.

## Computer Simulations.

To study the effect of adding chromatin loops to our model, we modified the Monte-Carlo simulations by Herrick *et al.*[8] in a number of ways. We accounted for the size of origin proteins ($\sim$10 nm) by using a lattice size $\Delta x$ = 116 basepairs (bp), which is fixed by setting the timestep of the simulation $\Delta t$ = 0.2 minutes ($\Delta x = v \cdot \Delta t$, where the fork velocity $v$ = 580 bp/min).[8] The parameters used in the simulation, such as the number and size of combed molecules, are the same as in the experiment, which justifies a direct comparison between the two.

The simulation consists of three stages: origin "licensing," "S phase," and "molecular combing." In the licensing stage, potential origins are distributed along each molecule (or lattice site). In the random-initiation scenario, the potential-origin sites are chosen at random from the unreplicated domains of DNA. In the loop-formation scenario explored below, they are chosen in a way that depends on the positions of the moving replication forks.

In the S phase stage, origins fire and forks grow bidirectionally, as in previous simulations. In the modified simulation incorporating the replication-factory model, there are multiple chromatin loops around each factory. Each potential origin has a different probability of initiation depending on how far it is from the two left and right approaching forks. To calculate the probability of loop formation for a single loop between a potential origin and the closest approaching fork, we used an approximation due to Ringrose et al.[23] (see also Eq. 3 in ref. [21]) that interpolates between the SY and Gaussian-chain distributions: $c \times l^{-3/2} \times \exp\left(-\frac{8}{l^2}\right)$, where $c$ is a normalization constant that should be determined based on the total number of new initiations at each time step, and $l$ the reduced length $L/l_p$. Note that the loop-formation probability is a function of the persistence length of the *Xenopus* chromatin, which has not been measured under the conditions applying to the present experiment. We fit an analytical approximation (Eq. 1) to the eye-to-eye distribution to obtain an estimate of the persistence length $l_p$.[22] We used the value from the fit (3.2 kb) in simulations incorporating the effects of loops. Then we determined how many origins to initiate, according to the experimentally determined initiation rate $I(t)$.[8] In each time step $\Delta t$, the number of initiations is $\Delta N(t) = I(t) \cdot \Delta t \cdot L'$, where $L'$ is the length of DNA that is unreplicated at time $t$, and the frequency of initiation $I(t)$ is the number of initiations per unit time per unit length, averaged over the genome. Once the probability of initiation for each potential origin and the $\Delta N(t)$ are determined, the corresponding number of potential origins is chosen for initiation by standard Monte-Carlo procedure (Fig. 2). In our computer program, we recorded only the positions of the forks themselves, rather than the state of every lattice site; this allowed us to carry out lengthy simulations (400-6300 runs; 20-200 Mb of DNA simulated in each run) using an ordinary desktop computer.

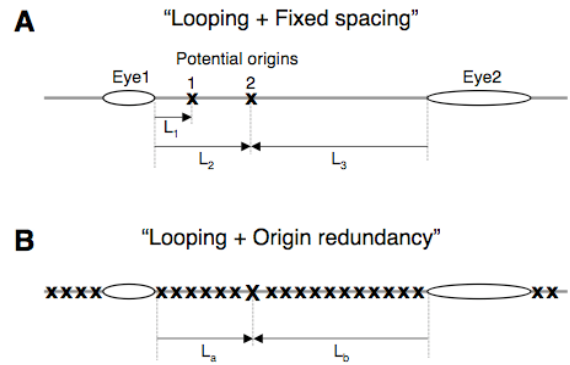

**Figure 2 - Computer simulation rules.** Initiation rules for the computer simulations. (A) Looping + fixed spacing: there are two replication bubbles and two potential origins (x) 1 and 2. The probability of initiation of each potential origin is $p_1$=SY($L_1$) and $p_2$=SY($L_2$), respectively, where SY(L) is the loop-formation probability (interpolated Shimada-Yamakawa distribution) of chromatin of loop-size L. Note that $p_2 \neq$ SY($L_3$) because $L_3 > L_2$. We first calculate $p$'s for all potential origins, and then we normalize the probabilities and initiate $\Delta N(t)$ potential origins using standard Monte Carlo procedure. (B) Looping + origin redundancy: initiation rules are the same as (A). Again, for potential origin X, the probability of initiation is SY(L), not SY($L_b > L_a$).

In the final molecular-combing stage, we cut the molecules into fragments whose size distribution matches that of the actual experiment (roughly Poissonian, with an average of 102 kb). We then coarse-grained the simulated molecules by averaging over a length scale of 480 bp (~ 0.24 µm) in order to account for the optical resolution of the experimental scanned images of combed molecules. The final result is a simulation of the experimental data set that includes the different biological scenarios of interest, in this case chromatin loop-formation. We applied exactly the same data analysis to the simulated data set as we did to the experimental data set.

## Results

In previous work,[8] we drew on basic observations of DNA replication

(1) DNA is organized into a sequential series of replication units, or replicons, each of which contains a single origin of replication;
(2) Each origin is activated not more than once during the cell-division cycle;
(3) DNA synthesis propagates at replication forks bidirectionally from each origin;
(4) DNA synthesis stops when two newly replicated regions of DNA meet;

to construct a "kinetic model" of DNA replication based on three assumptions:

(1) the initiation of origins could be described by a function $I(x,t)$ that gives the probability of initiating an origin at position $x$ along the genome at time $t$ during S phase;
(2) replicating domains expand symmetrically with a velocity $v$;
(3) replicating domains that impinge on each other coalesce.

We then used the mathematical model defined by these assumptions to analyze data from a recent experiment on DNA replication.[12] In this experiment, cell-free early-embryo *Xenopus* was dual-labelled with two fluorescent dyes. The first was present at the beginning of the replication cycle; the second was added at a controllable time point during S phase. DNA fragments were then isolated and combed onto substrates, where they were analyzed by two-color epifluorescence microscopy. The alternating patterns of labelling then gave a "snapshot" of the state of the DNA fragment at the time the second label was added. Statistical analysis of such labels gave empirical distributions of replicated domain ("eye") lengths, "hole" sizes between replicated lengths, and "eye-to-eye" distances, defined as the distance between the center of one eye and the center of a neighboring eye. From the averages of eyes, holes, and eye-to-eye lengths, we inferred the spatially averaged initiation rate $I(t)$, which is defined as the number of new initiations per unit time per unit unreplicated length, at time $t$.

Although the previous analysis successfully incorporated information deduced from the averages or the various distributions, we did not look at the distributions themselves. In particular, the eye-to-eye distribution is an important quantity in that it approximates the origin-spacing distribution for small eye-to-eye distances because both eyes involved must also be small and thus likely contain just one origin each. Here, we show that analysis of these quantities including neighborhood eye-size correlations lead us to refine the assumptions made in the kinetic model, shedding light on the long-standing random-completion problem in the process.

### The Eye-to-Eye Distribution Predicted Using Random Initiation Does Not Agree with Experiment.

We extracted the *distribution*, $\rho_{i2i}$, of distances separating centers of neighboring eyes (eye-to-eye distances) from the raw experimental data,[12] and compared it with the $\rho_{i2i}$ distribution obtained from a numerical simulation that assumed random distribution and activation of replication origins (data compiled from 6,300 runs of the simulation described in Herrick *et al.*[8]) (Fig. 3A).

The difference between the distributions, $\Delta\rho_{i2i} = \rho_{i2i\_exp} - \rho_{i2i\_random}$, is shown in Fig. 3B. Notice that there are two clearly distinct regimes. In the first regime ($l_{i2i} < \sim 20$ kb), the experimental data clearly differ from the simulation ($P=4\times10^{-33}$; $\chi^2=165$ for $n=6$ degrees of freedom).

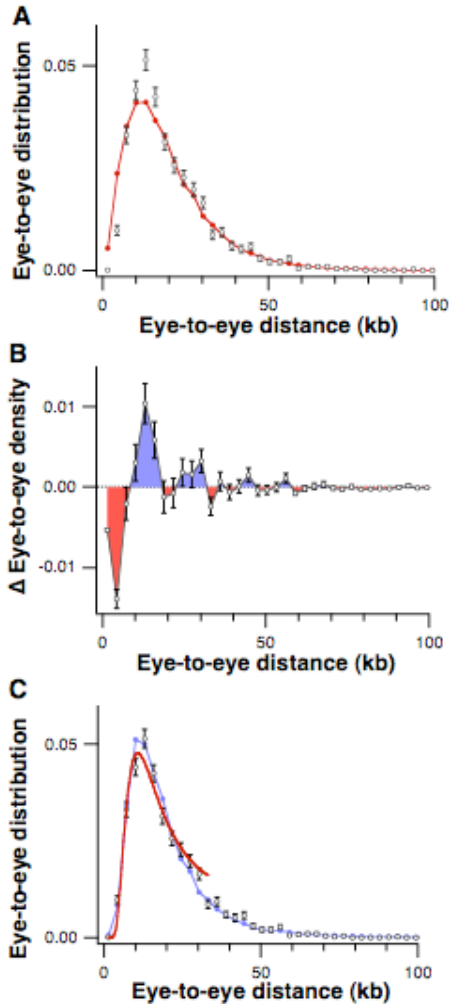

**Figure 3 - Distribution of replication origins and the loop-formation probability.** Because the shape of the eye-to-eye distribution changes little during most of *S*-phase, we pooled the experimental and simulation data for *f* = 10-90%, where *f* is the fraction of the genome that has been replicated. (A) Eye-to-eye distribution $\rho_{i2i}$. (O) Experiment; (●) Random initiation (simulation). (B) Difference between the experiment and assumed random initiations, $\Delta\rho_{i2i} = \rho_{i2i\_exp} - \rho_{i2i\_random}$. In the enhancement region (shaded blue above the zero line), more initiations occur than in the random case; in the exclusion zone (shaded red below the zero line), new initiations are inhibited. One can see that the first two oscillations ($l_{i2i} \leq 20$ kb) are statistically significant, while the agreement between $\rho_{i2i\_exp}$ and $\rho_{i2i\_random}$ becomes better as $l_{i2i}$ increases. (C) Experimental $\rho_{i2i}$ and the Shimada-Yamakawa loop-formation probability. The red curve is a fit to the Shimada-Yamakawa approximate distribution, Eq. 1, over the range 0-35 kb. The fit gives $l_p = 3.2 \pm 0.1$ kb. The fit value of persistence length is biased downwards slightly because the SY distribution becomes inaccurate beyond a few times the persistence length.[21] The blue curve is the result of a simulation incorporating loops of $l_p = 3.2$ kb, as discussed in the text.

Initiations are *inhibited* over origin-to-origin distances smaller than 8 kb (mostly smaller than 4-5 kb). This is consistent with both the observation that there is only one origin initiation event on plasmids smaller than ~10 kb[2] and the speculation that an exclusion zone ensures a minimum origin-to-origin distance.[7] On the other hand, activation of one origin appears to *stimulate* the activation of neighboring origins each separated by a distance of 8 – 16 kb (peak at ~ 13 kb). This number is consistent with the previously reported origin spacings of 5 – 15 kb[9,12] and the saturation density of *Xenopus* Origin Recognition Complexes (XORCs)[26,27] along sperm chromatin in egg extracts.

The second regime ($l_{i2i} \geq 20$ kb) shows that for simulation and experiment the distribution of large eye-to-eye distances is statistically similar ($P=0.14$; $\chi^2=34$ for $n = 26$), which implies that the random-initiation hypothesis holds for this regime, even as it fails at smaller origin separations.[28]

## Eye-Size Correlations and Origin Synchrony.

We next tested for the presence of correlations between the sizes of nearby eyes. Fig. 4 shows that there is a weak but statistically significant positive correlation: larger eyes tend to have larger neighbors, and *vice versa*. Because domains grow at constant velocity, size correlations may be interpreted as origin synchrony. The value for the nearest-neighbor correlation is consistent with that reported by Blow *et al*. (0.16).[9]

The observation of eye-size correlations has qualitative significance in that no local initiation function $I(x,t)$ – whatever its form – can produce correlations.[29] Intuitively, the presence of eye-size correlations means that the probability of initiating an origin is enhanced by the presence of nearby active origins and thus cannot be a function only of *x* and *t* (position along the genome and time during S phase). In Fig. 4, we calculate via Monte-Carlo simulation the eye-size correlations assuming that origins are placed at random along the genome (red points) and intiations are independent from one another. As expected, the correlations are consistent with zero.

**Origin Spacing, Loops, and Replication Factories.**
Since the experimental eye-to-eye distribution is not consistent with the random-initiation hypothesis for short distances (< 20 kb) and since eye-size correlations imply some kind of nonlocal interaction between origins, we tested an alternative hypothesis that chromatin folding can lead to a replication factory with loops (hereafter, the loop model),[16,17,30] against data. In the loop model, initiations occur at the replication factory, and there must be a correlation between the loop sizes and the distances between replication origins. As mentioned earlier, because of the intrinsic stiffness of chromatin, loops have a preferred size: activated origins will tend to occur at a characteristic separation from the replication forks of already activated replication origins.

In the Monte Carlo simulations, we compute for each time interval $\Delta t$ the number of initiations $\Delta N(t) = I(t) \cdot \Delta t \cdot L'$ (where $L'$ is the length of DNA that is unreplicated at time $t$, using the published result of $I(t)$,[8] that will occur throughout the genome. The distribution of $\Delta N(t)$ potential origins is not random but follows approximately the distribution of loop sizes predicted by Shimada and Yamakawa (SY) using a helical wormlike chain model of polymer (see Fig. 2).[22] In the SY model (see Materials and Methods), the distribution of loop sizes is peaked at 3-4 times the persistence length $l_p$ of the polymer. Thus, origins either too close or too far from the approaching forks have less probability of initiation than those a few times $l_p$ apart.

The results of our modified simulations are shown in Figs. 3C and 4 (data compiled from 400 runs of the simulation), which shows that incorporating the loop model into the initiation algorithm makes the $\rho_{i2i}$ data from the simulation agree with experiment. In Fig. 4, the simulation data (blue points) show eye-size correlations more consistent with experiment: this is an expected result since using the SY distribution, as a relative initiation probability of potential origins from approaching forks implicitly enforces clustering and rough synchrony of origin firings. In Fig. 3C, we plot both the SY and the measured $\rho_{i2i}$ distributions (red and blue curves, respectively). Note that the SY distribution itself should only approximate $\rho_{i2i}$ for the following reasons: The SY distribution gives the probability that the ends of a polymer meet, while the $\rho_{i2i}$ distribution gives the probability that two points along the DNA meet. Unlike the SY distribution, these points are constrained to be discrete loci along the DNA wherever there are potential origins. In addition, if a long loop containing additional potential origins forms, multiple loops may be created by subsequent binding of one of the potential-origin sites interior to the original loop. Such possibilities are not considered in the SY distribution. Still, for small loop sizes, neither of these effects is important because the high bending-energy cost inhibits subloop formation in loops that are already small, and we may compare the SY and $\rho_{i2i}$ distributions in this regime. The result, in Fig. 3C (red curve), is reasonably consistent with the data over the fit range (0-35 kb) and gives a persistence length of 3.2 ± 0.1 kb. This persistence length was then used for the simulation data (blue curve in Fig. 3C and blue points in Fig. 4). The optimal loop size is then ~ 11 kb (peak of curves in Fig. 3C), and the exclusion zone is approximately one persistence length, ~ 3-4 kb. These values are in excellent agreement with the observed average XORC saturation density 7-16 kb along the *Xenopus* sperm chromatin in egg extracts,[26,27] the known values of origin-spacings of 5-15 kb[9,12] and loop-sizes[31] of early embryo *Xenopus,* as well as the average origin-spacing 7.9 kb of transcriptionally quiescent *Drosophila* early embryos.[32]

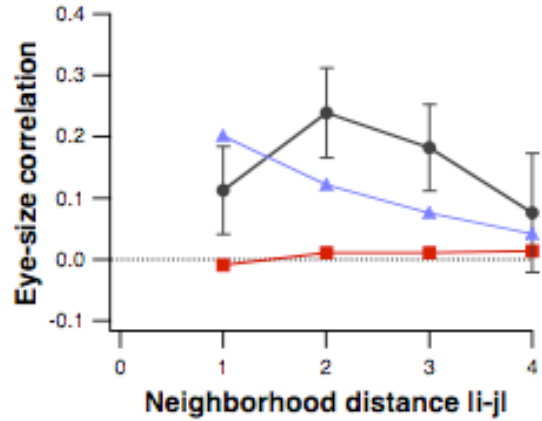

**Figure 4 - Eye-size correlation.** Eye-size correlation $C(|i-j|)$ vs. neighborhood distance $|i-j|$ between eyes for three different cases (data for $f$ = 40-60% pooled together): Experiment (●), random initiation (■),[8] and replication-factory model with loop-formation (▲) (each data set compiled from 400 runs of the simulation). The random-initiation case does not produce any correlations, as expected; however, both experiment and the replication-factory/loop-formation model produce statistically significant – and similar – positive correlations.

## Discussion

### Persistence Length.
The persistence length that we infer for *Xenopus* sperm chromatin fiber in egg extracts (3.2 ± 0.1 kb) is comparable to that found in other systems. Cui and Bustamante measured the persistence length of chromatin fiber under low-salt and in physiological conditions using force-extension curves obtained by stretching single chicken erythrocyte chromatin fibers.[19] They found $l_p$ = 30 nm, which corresponds to 3.5 kb for a typical packing ratio of 40,[21] slightly larger than our value. On the other hand, Dekker *et al.*[15] used their 3C technique to estimate $l_p$ for chromosome III in yeast, in the $G_1$ phase of its cell cycle. They found $l_p$ = 2.5 kb, slightly smaller than our value. Although these measurements are for different systems, their similarity suggests that chromatin stiffness may typically be in this range and also that the looping scenario examined here may apply more generally.

### The Random-Completion Problem.
As mentioned in the Introduction, because replication origins in embryos are not linked to sequence,[3] the relevant model of DNA replication must be able to address the random-completion problem –- i.e., it must be able to account both for the observed duration of S phase and the relative infrequency of long "fluctuations" of the time to copy the genome. The two scenarios discussed above – "origin redundancy" and "fixed spacing" – have issues of concern. One problem with the origin-redundancy scenario is that, until recently, potential origins were believed to be directly associated with XORCs by assembly of pre-replication complexes (pre-RCs) consisting of several proteins (XORC, CDC6, CDT1 and MCM2-7) before the start of S phase ("origin licensing").[1,3,6] The potential origins are then activated during S phase. The difficulty is that there are approximately the same number of XORCs as initiated origins. Recent data, however, suggest that all the MCM2-7 complexes, 10-40 of which are recruited by each XORC, may be competent to initiate replication and that the choice of MCM complex is not made before the start of S phase, implying that a much greater fraction of the genome serves as potential-origin sites.[33] Edwards *et al.* then showed that CDC45, which is essential for initiating replication at MCM complexes, is limiting for DNA replication, and, based on this observation, they further speculated that activation of the first MCM complexes may lead to inactivation of neighboring MCM complexes, thereby restricting initiation to defined intervals. Even so, restricting initiation itself does not prevent the formation of large gaps between origins, nor does it explain the significant eye-size correlations, i.e., partial synchrony in origin firings. In other words, one still needs a structural basis for regulation of origin spacing and origin synchrony.

The problem with the fixed-spacing scenario is its fragility: If even one origin fails to fire, the length of S phase would increase significantly (at least an order of 10 minutes for approximate XORC spacing 10 kb and fork velocity 600 bp/min).[6] Thus, this fixed-spacing scenario requires an unknown mechanism to ensure very high efficiency of origin initiation to prevent two or more nearest-neighbor origins from failing to initiate.

The loop model considered here incorporates elements of both scenarios. Like the origin-redundancy scenario, it is based on the measured, increasing $I(t)$. But the looping accounts naturally for the origin-exclusion zone, as well as the observation that individual origins may be more closely spaced than the typical exclusion-zone size. Like the fixed-spacing scenario, there is also regularity in the origin spacing. Here, that regularity appears as a natural consequence of the stiffness of chromatin, and no other mechanism is required. Both the redundant origins and the regularity contribute to making the failure to replicate the entire genome within the common duration of S phase unlikely.

In our case, we tested the loop model with various constraints on the distribution of potential-origin sites using computer simulations. The results shown here assumed an average potential-origin spacing of 7 kb, randomly distributed on a DNA molecule fragment whose length is approximately 500-1000 kb before being cut. The numbers reflect previously reported values for XORC spacings[26,27] and the average origin spacing.[9,12] The small size of the DNA fragments also prevents large gaps between origins, thus avoiding the random-completion problem. On the other hand, the assumption that MCM complexes completely cover the genome, and all are competent for initiation also produced a result that is similar to the one presented here when looping (and implicit synchrony rule) is incorporated in regulating initiation. At this point, the statistics available in the data of Herrick et al.[12] and the lack of theoretical understanding of chromatin behavior make it difficult to invert the data to draw conclusions about the form of the potential origin-site distribution. However, wide range of potential origin distributions considered above gave results consistent with an important biological role for chromatin looping.

We emphasize that the loop model not only gives a better quantitative explanation of the $\rho_{i2i}$ distributions, it also provides a basis for the correlations between neighboring eye sizes. Although the increase in initiation rate during S phase[7,8,12] can explain the observed duration of genome replication, it cannot give rise to correlations on its own. Some mechanism wherein the initiation of one origin has effects on the likelihood of nearby initiations is required. The detailed analysis of the experimental data presented here shows that inhibition near activated origins, coupled with enhancement at a characteristic farther distance is required. We argue that loops are the simplest, most natural mechanism that can satisfy these requirements.

**Chromatin Loops and Replication Kinetics.**
Our findings imply that higher-order chromatin structure may be tightly linked to the kinetics of DNA replication in the early-embryo *Xenopus laevis* in-vitro system. We note that looping is a well-established way for DNA-bound proteins to interact over long distances.[34] At scales of hundreds of bases, it plays an important role in gene regulation. For example, the looping of dsDNA ($l_p$=150 bp) with intrinsic curvature facilitates greatly the interaction between regulatory proteins at upstream elements and the promoter.[35] Loops are also known to appear in higher-order chromatin structures such as the 30-nm fiber at scales of thousands of bases, or even longer.[36] For example, Buonguorno-Nardelli et al.[31] established a correlation between chromosomal loop sizes and the size of replicated domains emanating from a single replication origin (replicon). Chromatin loops are also a natural part of the replication factory model of DNA replication, where polymerases and their associated proteins are localized in discrete foci, with chromosomes bound to the factory complex at multiple nearby points along the genome.[16,17]

The natural follow-up to the results presented here would be to assess the generality of our results: Do they extend to other early-embryo systems? Are they valid in vivo? Do they apply to other transcriptionally quiescent regions of the genome?

Based on our results, we can also predict how altering chromatin structure should affect DNA replication. For example, if the replication factory model is correct, the loop size is roughly the origin spacing. Since the optimal loop size is proportional to $l_p$, the duration of S phase increases with $l_p$ in a way that can be modeled quantitatively using the simulation. One experimental approach to testing these ideas would be to combine combing and single-molecule elasticity experiments on *Xenopus*, isolating DNA from different regions of the genome. If there is heterogeneity in the stiffness of chromatin fibers in the genome, we would predict a corresponding heterogeneity in the origin-spacing distribution.

**Loop Formation and Replication Factories.**

Currently, there are no direct experimental observations of the internal structure of replication factories. For example, the number of replicons or loops per individual factories or foci is only estimated indirectly from various quantities such as total number of origins, number of foci, fork velocities, and rough origin spacing. However, replication foci appears to be universal features of eukaryotic DNA replication and nuclear structure, and in mammalian cells, they are globally stable structures, with constant dimensions, that persist during all cell cycle stages including mitosis (for a review see ref. [30]). On the other hand, experimental evidence suggests that chromatin is very dynamic within individual foci at the molecular level (see, for example, ref. [37]), consistent with our computer simulations.

In Fig.1, a schematic diagram shows how chromatin folding can lead to a replication factory with loops. Once loops form, they can dynamically fluctuate locally around factories throughout interphases, with highest mobility during $G_1$ phase, while the global structures of foci are stable within nucleus. We note that recent theoretical calculations show that such chromatin folding can be very fast ($10^{-3}$-$10^{-2}$ sec), and the loop-formation time is inversely proportional to the SY distribution. In other words, loop-formation is fastest when its size is 3-4 times the persistence length, and it increases exponentially as the loop size becomes smaller than the persistence length (see Eq.1 in Materials and Methods), leading us to further speculate that the origin-spacing in *Xenopus* or *Drosophila* early embryos may be selected to maximize the loop-formation and contact rate of origins.

On the other hand, the exact physical mechanisms of initiation and its partial synchrony within individual replication factory remain for future experiments. For example, although the eye-size correlation in our simulation decreases monotonically, the experimental data do not rule out the possibility of non-monotonic decay. Also, the correlations from both simulation and experiment are significant but weak. This suggests that the synchrony within a replication factory is not perfect, and nearest neighbor origins do not necessarily fire simultaneously.[30]

Regardless of the biological complexity in replication foci, however, we emphasize that the loop sizes are determined by the basic physical principles explained above, namely, the balance between chromatin energy and entropy.

## Conclusion

In *Xenopus* early embryos, replication origins do not require any specific DNA sequences, but the whole genome (3 billion bases) is completely duplicated within 10-20 minutes. This implies that there must be a mechanism that regulates replication other than sequence in this system.

The results presented here provide strong evidence that a combination of redundant origins and chromatin loops together provide such a mechanism. We find that the persistence length of chromatin loops plays a biological role in DNA replication, in that it determines the optimal distances between replication origins in *Xenopus* early embryos. Chromatin loops constitute a structural basis for the observed distribution of replication origins in *Xenopus* early embryos, accounting for both origin exclusion zones and origin clustering along the genome. It would also be interesting to see whether the same scenario applies to other early-embryo systems such as *Drosophila*.

The picture of the replication process presented here also leads naturally to more detailed hypotheses about the role of chromatin, which should stimulate further modeling efforts.

Finally, it would be highly desirable to vary the persistence length of chromatin, to see whether the origin spacings change in a way predicted by our theory. Although such an experiment poses formidable challenges, it would be an important step forward in understanding the role of chromatin structure in DNA replication.

## Acknowledgements
We thank G. Almouzni, B. Arcangioli, B.-Y. Ha, A. Libchaber, J. Marko, B. Sclavi, and E. Svetlova for critical reading of the manuscript. We are particularly grateful to J.A.Huberman and P. Pasero for their numerous and invaluable comments and discussions. This work was supported by NSERC (Canada), la Fondation de France, and NIH.

## References

1. Gilbert DM. Making sense of eukaryotic DNA replication origins. Science 2001; 294: 96-100.
2. Hyrien O, Méchali M. Chromosomal replication initiates and terminates at random sequences but at regular intervals in the ribosomal DNA of *Xenopus* early embryos. EMBO J 1993; 12:4511-4520.
3. Blow JJ. Control of chromosomal DNA replication in the early *Xenopus* embryo. EMBO J 2001; 20:3293-3297.
4. Raff JW, Glover DM. Nuclear and cytoplasmic mitotic cycles continue in *Drosophila* embryos in which DNA synthesis is inhibited with aphidicolin. J Cell Biol 1988; 48:399-407.
5. Laskey RA: Chromosome replication in early development of *Xenopus laevis*. J Embryol Exp Morphol 1985; 89:285-296.
6. Hyrien O, Marheineke K, Goldar A. Paradoxes of eukaryotic DNA replication: MCM proteins and the random completion problem. BioEssays 2003; 25:116-125.
7. Lucas I, Chevrier-Miller M, Sogo JM, Hyrien O. Mechanisms ensuring rapid and complete DNA replication despite random initiation in *Xenopus* early embryos. J Mol Biol 2000; 296:769-786.
8. Herrick J, Jun S, Bechhoefer J, Bensimon A. Kinetic model of DNA replication in eukaryotic organisms. J Mol Biol 2002; 320:741-750.
9. Blow JJ, Gillespie PJ, Francis D, Jackson DA. Replication origins in *Xenopus* egg extract Are 5-15 kilobases

apart and are activated in clusters that fire at different times. J Cell Biol 2001; 152:15-25.
10. Bensimon A, Simon A, Chiffaudel A, Croquette V, Heslot F, Bensimon D. Alignment and sensitive detection of DNA by a moving interface. Science 1994; 265:2096-2098.
11. Norio P, Schildkraut CL. Visualization of DNA replication on individual *Epstein-Barr* virus episomes. Science 2001; 294:2361-2364.
12. Herrick J, Stanislawski P, Hyrien O, Bensimon A. Replication fork density increases during DNA synthesis in *X. laevis* egg extracts. J Mol Biol 2000; 300:1133-1142.
13. Raghuraman MK, Winzeler EA, Collingwood D, Hunt S, Wodicka L, Conway A *et al.* Replication dynamics of the yeast genome. Science 2001; 294:115-121.
14. Wyrick JJ, Aparicio JG, Chen T, Barnett JD, Jennings EG, Young RA, Bell SP, Aparicio OM. Genome-wide distribution of ORC and MCM proteins in *S. cerevisiae*: high-resolution mapping of replication origins. Science 2001; 294:2357-2360.
15. Dekker J, Rippe K, Dekker M, Kleckner N. Capturing chromosome conformation. Science 2002; 295:1306-1311.
16. Cook PR. The organization of replication and transcription. Science 1999; 284:1790-1795.
17. Méchali M. DNA replication origins: from sequence specificity to epigenetics. Nat Rev Genet 2001; 2:640-645.
18. Yamakawa H. Helical wormlike chains in polymer solutions. Berlin: Springer, 1997.
19. Cui Y, Bustamante C. Pulling a single chromatin fiber reveals the forces that maintain its higher-order structure. Proc Natl Acad Sci 2000; 97:127-132.
20. Allemand JF, Bensimon D, Jullien L, Bensimon A, and Croquette V. pH-dependent specific binding and combing of DNA. Biophys J 1997; 73:2064-2070.
21. Rippe K. Making contacts on a nucleic acid polymer. Trends Biochem Sci 2001; 26:733-740.
22. Shimada J, Yamakawa H. Ring-closure probabilities for twisted wormlike chains. Application to DNA. Macromolecules 1984; 17:689-698.
23. Ringrose L, Chabanis S, Angrand P, Woodroofe C, Stewart AF. Quantitative comparison of DNA looping in vitro and in vivo: chromatin increases effective DNA flexibility at short distances. EMBO J 1999; 18:6630-6641.
24. Jun S, Bechhoefer J, Ha BY. Diffusion-limited loop formation of semiflexible polymers: Kramers theory and the intertwined time scales of chain relaxation and closing. Europhys Lett 2003; 64 (3): 420-426.
25. Merlitz H, Rippe K, Klenin KV, Langowski J. Looping dynamics of linear DNA molecules and the effect of DNA curvature: A study by Brownian Dyanmics simulation. Biophys J 1998; 74: 773-779.
26. Rowles A, Chong JP, Brown L, Howell M, Evan GI, Blow JJ. Interaction between the origin recognition complex and the replication licensing system in *Xenopus*. Cell 1996; 87:287-296.
27. Walter J, Newport JW. Regulation of replicon size in *Xenopus* egg extracts. Science 1997; 275:993-995.
28. The agreement between the two curves (experiment and random initiation) becomes better as the eye-to-eye distance increases. However, we note that the *P*-values for the two regimes (inhibition and enhancement) are most distinguishable when they are divided after the first two oscillations, i.e. at around $l_{i2i}$ = 20 kb. On the other hand, we also note that the simple random-initiation hypothesis reproduces all mean quantities such as the mean eye size throughout S phase very well, as shown in [8].

29. Sekimoto K. Evolution of the domain structure during the nucleation-and-growth process with non-conserved order parameter. Int J Mod Phys B 1991; 5:1843-1869.
30. Berezney R, Dubey DD, Huberman JA. Heterogeneity of eukaryotic replicons, replicon clusters, and replication foci. Chromosoma 2000; 108:471-484.
31. Buongiorno-Nardelli M, Micheli G, Carri MT, Marilley M. A relationship between replicon size and supercoiled loop domains in the eukaryotic genome. Nature 1982; 298:100-102.
32. Blumenthal AB, Kriegstein HJ, Hogness DS. The units of DNA replication in *Drosophila melanogaster* chromosomes. Cold Spring Harbor Symp Quant Biol 1973; 38:205-223.
33. Edwards MC, Tutter AV, Cvetic C, Gilbert CH, Prokhorova TA, Walter JC. MCM2-7 complexes bind chromatin in a distributed pattern surrounding the origin recognition complex in *Xenopus* egg extracts. J Biol Chem 2002; 277:33049-33057.
34. Schleif R. DNA looping. Annu Rev Biochem 1992; 61:199-223.
35. Rippe K, von Hippel PH, Langowski J. Action at a distance: DNA-looping and initiation of transcription. TIBS 1995; 20:500-506.
36. Horn PJ, Peterson CL. Molecular biology. Chromatin higher order folding--wrapping up transcription. Science 2002; 297:1824-1827.
37. Heun P, Laroche T, Shimada K, Furrer P, Gasser SM. Chromosome Dynamics in the Yeast Interphase Nucleus. Science 2001; 294:2181-2186.